\documentclass[twocolumn,english,superscriptaddress]{revtex4-1}
\usepackage[T1]{fontenc}
\usepackage[latin9]{inputenc}
\usepackage{verbatim}
\usepackage{amsbsy}
\usepackage{amstext}
\usepackage{graphicx}

\makeatletter
\usepackage{babel}

\makeatother

\usepackage{babel}
\begin{document}
\title{Holographic spin alignment for vector mesons}
\author{Xin-Li Sheng}
\email{sheng@fi.infn.it}

\affiliation{INFN Sezione di Firenze, Via G. Sansone 1, I-50019, Sesto F.no (Firenze),
Italy}
\affiliation{Institute of Particle Physics and Key Laboratory of Quark and Lepton Physics (MOS), \\Central China Normal University, Wuhan 430079, China}

\author{Yan-Qing Zhao}
\email{zhaoyanqing@mails.ccnu.edu.cn}
\affiliation{Institute of Particle Physics and Key Laboratory of Quark and Lepton Physics (MOS), \\Central China Normal University, Wuhan 430079, China}

\author{Si-Wen Li}
\email{siwenli@dlmu.edu.cn}
\affiliation{Department of Physics, College of Science, Dalian Maritime University, Dalian 116026, China}

\author{Francesco Becattini}
\email{becattini@fi.infn.it}
\affiliation{Department of Physics, University of Florence and INFN Via G. Sansone 1, I-50019, Sesto F.no (Firenze), Italy}

\author{Defu Hou}
\email{houdf@mail.ccnu.edu.cn}
\affiliation{Institute of Particle Physics and Key Laboratory of Quark and Lepton Physics (MOS), \\Central China Normal University, Wuhan 430079, China}

\begin{abstract}
We develop a general framework for studying the spin alignment $\rho_{00}$ for flavorless vector mesons by using the gauge/gravity duality. Focusing on the dilepton production through vector meson decay, we derive the relation between production rates at each spin channel and meson's spectral function, which can be evaluated by holographic models for a strongly coupled system. As examples, we study  $\rho_{00}$ for $J/\psi$ and $\phi$ mesons, induced by the relative motion to a thermal background, within the soft-wall model. We show that $\rho_{00}$ in the helicity frame for $J/\psi$ and $\phi$ mesons have positive and negative deviations from 1/3 at $T=150$ MeV, respectively, which consequently leads to different properties for their global spin alignments.
\end{abstract}
\maketitle

\section{Introduction}

\label{sec:00_intro} %%%	
Relativistic heavy-ion collisions provide a unique opportunity to test our understanding about the strongly interacting matter. In non-central collisions, the incoming nuclei carry huge orbital angular momentum (OAM), which can partly transfer to the Quark Gluon Plasma in the fireball and consequently induce the polarization of quarks or hadrons. A typical example is the global spin polarization for $\Lambda$ hyperons, observed by the STAR collaboration in Au+Au collisions \cite{Liang:2004ph,Voloshin:2004ha,Becattini:2007sr,STAR:2017ckg,STAR:2018gyt} (see e.g. \cite{Becattini:2020ngo,Gao:2020lxh,Huang:2020dtn,Becattini:2022zvf}, for recent reviews). As spin-one particles,
vector mesons should also be globally polarized by the OAM.

The most successful model quantitatively reproducing $\Lambda$ polarization is the hydrodynamic-statistical
model \cite{Becattini:2007sr,Becattini:2013fla} which relates spin polarization to the gradient of the
hydrodynamic fields (specifically, thermal vorticity and thermal shear, see ref.\cite{Becattini:2022zvf}).
However, at least up to non-dissipative terms, this model predicts a spin alignment of
vector mesons which is consistently lower than that measured by the experiments \cite{ALICE:2019aid,STAR:2022fan}. More models have been proposed to account for the observed
spin alignment. A model based on recombination of polarized quarks \cite{Liang:2004xn,Yang:2017sdk}
also leads to somewhat too low values. Another model was proposed in Refs. \cite{Sheng:2019kmk,Sheng:2020ghv,Sheng:2022wsy,Sheng:2022ffb,Sheng:2023urn}
where it is argued that strong force field fluctuations are responsible for the vector meson's spin
alignment; while these fluctuations does not contribute to the polarization of hyperons due to its
vanishing mean value, they generate a spin alignment of flavor singlet vector mesons through the
correlation of spin polarization of the constituent quark and antiquark \cite{Sheng:2022wsy,Sheng:2022ffb}.
Other possible mechanisms for the generation of spin alignment have been proposed in
\cite{Xia:2020tyd,Gao:2021rom,Muller:2021hpe,Kumar:2023ghs,Li:2022vmb,Sheng:2022ssp,Wei:2023pdf,Dong:2023cng,Fang:2023bbw}, which are still waiting for a quantitative verification.

\begin{comment}
Noncentral relativistic heavy-ion collisions provide an opportunity to investigate the interplay between the global orbital angular momentum (OAM) and the spin of elementary particles in the quark-gluon plasma, as demonstrated by the Barnett effect \cite{Barnett:1935wyv} and the Einstein-de Hass effect \cite{Einstein1915deutsche}. A typical example is the global spin polarization of the strange quark, which is transferred to the polarization of $\Lambda$ hyperons observed by the STAR collaboration in Au+Au collisions \cite{Liang:2004ph,Voloshin:2004ha,Becattini:2007sr,STAR:2017ckg,STAR:2018gyt} (see e.g. \cite{Becattini:2020ngo,Gao:2020lxh,Huang:2020dtn,Becattini:2022zvf}, for recent reviews).
\end{comment}

The spin state of a vector meson is described by a $3\times3$ spin density matrix $\rho_{\lambda_1 \lambda_2}$, with subscripts $\lambda_1,\lambda_2=0,\pm 1$ denote the spin quantum number along the quantization direction.
Due to the parity conservation in strong and electromagnetic interactions, the spin polarization of vector
meson, proportional to $\rho_{11}-\rho_{-1,-1}$, cannot be measured by the angular distribution of their
decay products. Instead, it is possible to measure the spin alignment $\rho_{00}$, denoting the probability
for the state with $\lambda=0$. In experiments, this is achieved through the polar angle distribution of
decay products in either p-wave strong decays such as $\phi\rightarrow K^+ +K^-$ \cite{Liang:2004xn,Yang:2017sdk,ALICE:2019aid,STAR:2022fan}, or dilepton decays such as $J/\psi\rightarrow \mu^++\mu^-$ \cite{Schilling:1969um,Faccioli:2010kd,ALICE:2020iev,ALICE:2022dyy}. For the later process, the polar angle distribution $W(\theta)$ is usually parameterized by $\lambda_\theta$ as \cite{Faccioli:2010kd}
\begin{equation}
W(\theta)\propto\frac{1}{3+\lambda_{\theta}}(1+\lambda_{\theta}\text{cos}^{2}\theta),\label{ang_dist}
\end{equation}
where $\theta$ is the angle between the momentum of one daughter particle and the chosen quantization direction, determined in the rest frame of meson. For dilepton decays, the $\lambda_\theta$ parameter is related to $\rho_{00}$ through the following relation,
\begin{equation}
\lambda_{\theta}=\frac{1-3\rho_{00}}{1+\rho_{00}}.\label{lambda_theta}
\end{equation}
Particularly, the two cases with $\lambda_\theta=-1$ (corresponding to $\rho_{00}=1$) and $\lambda_\theta=1$ ($\rho_{00}=0$), are called longitudinally and transversely polarized, respectively.  Recent measurements by the STAR collaboration at RHIC beam-energy-scan energies \cite{STAR:2022fan} and by the ALICE collaboration at the LHC energy \cite{ALICE:2022dyy} show that the $\phi$ and $J/\psi$ mesons are preferably to be longitudinally and transversely polarized respectively. However, there are no currently theoretical works to understand the difference between $\rho_{00}$ for $\phi$ and $J/\psi$ systematically and consistently, thus this motivates us to provide a framework in theory to fill this blank, in particular, by using AdS/CFT correspondence and holography since it may relate probably to the non-perturbative property of hadron.

Indeed, the most essential reason to use AdS/CFT and holography to investigate spin alignment is due to its powerful method to analyze the strongly coupled system of QFT \cite{Maldacena:1997re,Aharony:1999ti,Witten:1998qj,Witten:1998zw}. Specifically, the information of a strongly coupled QFT system can be obtained by perturbing its dual classical gravity theory in holography. Since QCD is a typical strongly-coupled QFT, it has attracted great interest to study hadron physics and QCD in strong-coupled region through AdS/CFT and holography for a very long time, e.g. top-down approaches based on D-brane, gauge-gravity duality, and string theory in \cite{Babington:2003vm,Kruczenski:2003uq,Sakai:2004cn,Sakai:2005yt,Li:2015uea,Li:2016gtz,Li:2022hka,Li:2023iuf,Li:2022wwv} or bottom-up approaches in \cite{Gubser:2008yx,Gursoy:2008bu}. Remarkably, the prescription to compute the two-point retarded Green functions in AdS/CFT, which has been justified in many different ways e.g.\cite{Gubser:1998bc,Son:2002sd,Policastro:2002se,Freedman:1998tz,vanRees:2009rw,Iqbal:2008by,Marolf:2004fy,Iqbal:2009fd,Lee:2008xf,Casalderrey-Solana:2011dxg,Liu:2009dm,Li:2023wyb}, is the key part to evaluate the spin alignment. As we will see, once the holographic action for meson is obtained on the gravity side, it is possible to employ the standard method in AdS/CFT to compute the two-point Green function of the dual mesonic current to evaluate the spin alignment. So in this work, we discuss first the generic form of the Green function associated with the holographic action for meson, then without loss of generality test our framework with the QCD soft-wall model and finally attempt to fit the data to understand various experimental effects related to spin alignment.

The outline of this letter is as follows. In Sec. \ref{sec:dilepton}, we focus on the dilepton decay for a flavorless vector meson and build the relation between observed spin alignment and the meson's spectral function. Then we provide a general framework in Sec. \ref{sec:general_spectral} for calculating the spectral function with holographic models. As enlightening examples, in Sec. \ref{sec:examples} we apply the soft-wall model to $J/\psi$ and $\phi$ mesons and discuss the behaviours of their spin alignment. We finally summarize this paper in Sec. \ref{sec:summary}. Throughout this paper, we use $\eta_{\mu\nu}=\eta^{\mu\nu}=\text{diag}(-1,1,1,1)$ for the metric in Minkowski space, and $g_{\mu\nu}$ for a generic metric in curved spacetime.

\section{Dilepton production through vector meson decay}\label{sec:dilepton}

We first consider the dilepton production through the decay of a flavor singlet
vector meson, such as $J/\psi\rightarrow l+\overline{l}$ and $\phi\rightarrow l+\overline{l}$.
The S-matrix element for the process of an initial state $i$ to a
final state $f$ plus a lepton pair $l\overline{l}$ is \cite{Gale:1990pn},
\begin{equation}
S_{fi}=\int d^{4}xd^{4}y\left\langle f,l\overline{l}\left|J_{\mu}(y)G_{R}^{\mu\nu}(x-y)J_{\nu}^{l}(x)\right|i\right\rangle \label{matrix_element}
\end{equation}
Phenomenologically, Eq. \ref{matrix_element} describes a two-step
process: 1) the vector meson is produced at the spacetime point $y$,
and 2) the vector meson decays into dileptons at point $x$. Here $G_{R}^{\mu\nu}$
is the retarded propagator for the vector meson which vanishes for
$x^{0}-y^{0}<0$ and therefore ensures the causality. In Eq. \ref{matrix_element},
$J_{\nu}$ is the current that couples to the vector meson.
It denotes the quark current (hardonic current) when we consider the
quark-gluon plasma (hardon gas) as the initial state. On the other
hand, $J_{\mu}^{l}$ denotes the leptonic current
\begin{equation}
J_{\nu}^{l}(x)=g_{Ml\overline{l}}\overline{\psi}_{l}(x)\Gamma_{\nu}\psi_{l}(x)
\end{equation}
where $\Gamma_{\mu}$ is the effective vertex for the vector meson-dilepton
interaction, $g_{Ml\overline{l}}$ is the coupling strength, and $\psi_{l}$
is the lepton field. In general, $\Gamma_{\mu}$ could be linear combinations
of all possible $4\times4$ matrices that behaves like Lorentz vectors,
such as $\overrightarrow{\partial}^{\mu}\pm\overleftarrow{\partial}^{\mu}$,
$\gamma^{\mu}$, $\sigma^{\mu\nu}(\overrightarrow{\partial}_{\nu}\pm\overleftarrow{\partial}_{\nu})$,
etc.
For simplicity, we only consider $\Gamma_{\nu}\approx\gamma_{\nu}$
in this paper. The transition probability for the process in Eq.
(\ref{matrix_element}) is given by $R_{fi}\equiv|S_{fi}|^{2}/(TV)$,
where $TV$ is the spacetime volume of the considered system, which
is assumed to be sufficiently large. We then sum over the final states
and average for the initial state over a thermal bath. Afterwards, we arrive
at the differential production rate $n(x,p)\equiv dN/(d^{4}xd^{4}p)$
for dileptons as a function of total momentum $p^{\mu}=(\omega,{\bf p})$,
\begin{eqnarray}
n(x,p) & = & -\frac{2g_{Ml\overline{l}}^{2}}{3(2\pi)^{5}}\left(1-\frac{2m_{l}^{2}}{p^{2}}\right)\sqrt{1+\frac{4m_{l}^{2}}{p^{2}}}p^{2}n_{B}(\omega)\nonumber \\
 &  & \times\left(\eta_{\mu\nu}+\frac{p_{\mu}p_{\nu}}{p^{2}}\right)G_{A}^{\mu\alpha}(p)\varrho_{\alpha\beta}(p)G_{R}^{\beta\nu}(p)\label{dilepton_vector}
\end{eqnarray}
where $G_{A}^{\mu\nu}$ is the advanced propagator, $n_{B}(\omega)=1/(e^{\omega/T}-1)$
is the Bose-Einstein distribution, and
\begin{equation}
\varrho_{\alpha\beta}(p)\equiv-\text{Im}D_{\alpha\beta}(p)\label{eq:spectra-funtion}
\end{equation}
is defined with the imaginary part of the retarded current-current
correlation
\begin{equation}
D^{\mu\nu}(p)\equiv\int d^{4}y\ \theta(y^{0})\left\langle \left[J^{\mu}(y),J^{\nu}(0)\right]\right\rangle e^{-ip\cdot y}\label{eq:current-current-correlation}
\end{equation}
Here $\left\langle O\right\rangle $ denotes the thermodynamical
average at a given temperature.

We emphasize that $\varrho_{\alpha\beta}$ in Eq. (\ref{eq:spectra-funtion})
is interpreted as the meson's spectral function in the thermal medium,
while $G_{R/A}^{\mu\nu}$ in Eq. (\ref{dilepton_vector}) are propagators
in vacuum, describing the propagating of vector meson after the freeze-out.
Eq. (\ref{dilepton_vector}) reduce to the dilepton production
rates through photon decay \cite{Burnier:2008ia,McLerran:1984ay,Weldon:1990iw,Gale:1990pn}
if we replace $G_{R/A}^{\mu\nu}$ by the vacuum propagator for a virtual
photon, $\eta^{\mu\nu}/p^{2}$. For the dilepton production through vector
meson decay, the propagators take the following form
\begin{equation}
G_{R/A}^{\mu\nu}=-\frac{\eta^{\mu\nu}+p^{\mu}p^{\nu}/p^{2}}{p^{2}+m_{V}^{2}\pm im_{V}\Gamma}
\end{equation}
where $m_{V}$ is the vacuum mass and $\Gamma$ is the width. On the
other hand, the spectral function can be decomposed using a complete
basis of polarization vectors
\begin{equation}
\varrho^{\mu\nu}(p)=\sum_{\lambda,\lambda^{\prime}=0,\pm1}v^{\mu}(\lambda,p)v^{\ast\nu}(\lambda^{\prime},p)\widetilde{\varrho}_{\lambda\lambda^{\prime}}(p)\,,\label{spectral_spin_space-1}
\end{equation}
where the polarization vectors
\begin{equation}
v^{\mu}(\lambda,p)=\left(\frac{{\bf p}\cdot\boldsymbol{\epsilon}_{\lambda}}{M},\boldsymbol{\epsilon}_{\lambda}+\frac{{\bf p}\cdot\boldsymbol{\epsilon}_{\lambda}}{M(\omega+M)}{\bf p}\right)\label{polarization_vectors}
\end{equation}
satisfy the orthonormality conditions $\eta_{\mu\nu}v^{\mu}(\lambda,p)v^{\ast\nu}(\lambda^{\prime},p)=\delta_{\lambda\lambda^{\prime}}$
and the completeness condition $\sum_{\lambda}v^{\mu}(\lambda,p)v^{\ast\nu}(\lambda,p)=(\eta^{\mu\nu}+p^{\mu}p^{\nu}/p^{2})$.
Here $M\equiv\sqrt{\omega^{2}-{\bf p}^{2}}$ is the invariant mass
for the vector meson. The three-component vectors $\boldsymbol{\epsilon}_{\lambda}$
denotes the spin direction in the meson's rest frame, with $\boldsymbol{\epsilon}_{0}$
being the spin quantization direction and $\boldsymbol{\epsilon}_{\pm1}$
being orthogonal to $\boldsymbol{\epsilon}_{0}$. If the quantization
direction is taken as the direction of ${\bf p}$ , one can find that
$\sum_{\lambda=\pm}v^{\mu}(\lambda,p)v^{\ast\nu}(\lambda,p)$ and
$v^{\mu}(0,p)v^{\ast\nu}(0,p)$ reduce to the transverse and longitudinal
projection operators in textbooks \cite{kapusta1989finite}, respectively. Then we are able
to separate contributions from different spin states $\lambda=0,\pm$1
to the dilepton production rate in Eq. (\ref{dilepton_vector}),
\begin{eqnarray}
n_{\lambda}(x,p)&=&-\frac{2g_{Ml\overline{l}}^{2}}{3(2\pi)^{5}}\left(1-\frac{2m_{l}^{2}}{p^{2}}\right) \nonumber\\
&&\times\sqrt{1+\frac{4m_{l}^{2}}{p^{2}}}\frac{p^{2}n_{B}(\omega)\widetilde{\varrho}_{\lambda\lambda}(p)}{(p^{2}+m_{V}^{2})^{2}+m_{V}^{2}\Gamma^{2}} \label{eq:dilepton-lambda}
\end{eqnarray}
The total dilepton number is the summation of all spin states, $n=\sum_{\lambda=0,\pm1}n_{\lambda}$.
The spin alignment for the vector meson resonance is defined as the
probability of the spin-zero state,
\begin{equation} \label{rho00}
\rho_{00}(x,{\bf p})\equiv\frac{\int d\omega\ n_{0}(x,p)}{\sum_{\lambda=0,\pm1}\int d\omega\ n_{\lambda}(x,p)}
\end{equation}
Since $n_{\lambda}$ is positive definite, the spin alignment is restricted
in the region $0\leq\rho_{00}\leq1$. We note that in general the width $\Gamma$ is much smaller than meson's vacuum mass $m_V$, thus Eq. (\ref{eq:dilepton-lambda}) behaves as a narrow peak around $-p^2=m_V^2$ and therefore the spin alignment is approximate to the ratio $\widetilde{\varrho}_{00}/\sum_\lambda\widetilde{\varrho}_{\lambda\lambda}$ at energy $\omega=\sqrt{m_V^2+{\bf p}^2}$.

\section{General setup for mesonic correlation function} \label{sec:general_spectral}

Since the key part of evaluating the spin alignment is to compute the correlation function as it is discussed in the last section, let us outline the general setup for the correlation function of vector meson by using AdS/CFT in this section. The AdS/CFT dictionary,
\begin{equation}
Z_\text{QFT}\left[A_{\mu}^{\left(0\right)}\right]=Z_\text{gravity}\left[A_{\mu}\right],
\end{equation}
where
\begin{eqnarray}
Z_\text{QFT}\left[A_{\mu}^{\left(0\right)}\right]&=&\left\langle \exp\left\{ \int_{\partial\mathcal{M}}J^{\mu}A_{\mu}^{\left(0\right)}d^{4}x\right\} \right\rangle ,\nonumber\\Z_\text{gravity}\left[A_{\mu}\right]&=&\exp\left\{ -S_\text{bulk}\left[A_{\mu}\right]\right\} , \label{adsdic}
\end{eqnarray}
illustrates that the generating functional $Z_{\mathrm{QFT}}$ of the dual QFT living at boundary $\partial\mathcal{M}$ is equivalent to the partition function $Z_{\mathrm{Gravity}}$ of the bulk gravity on $\mathcal{M}$. Here $A_{\mu}$ is a bulk vector on $\mathcal{M}$ and $A_{\mu}^{\left(0\right)}$ is its boundary value which is recognized as the source of operator $J^{\mu}$ on $\partial\mathcal{M}$, i.e. $A_{\mu}^{\left(0\right)}=A_{\mu}|_{\partial\mathcal{M}}$. We note that the relation presented in Eq. \ref{adsdic} is workable only if the dual QFT is a strongly coupled theory. In general, the bulk geometry should be AdS-like which takes the following form of the metric as,
\begin{equation}
ds^{2}=g_{\mu\nu}dx^{\mu}dx^{\nu}+g_{\zeta\zeta}d\zeta^{2}\,,\label{metric}
\end{equation}
where $g_{\mu\nu}, g_{\zeta\zeta}$ only depend on the holographic radial coordinate $\zeta$ and the indices $\mu,\nu=0,..3$ label the four-dimensional spacetime. The coordinate $\zeta$ is defined in the region $0\leq\zeta\leq\zeta_{h}$, where $\zeta=0$ is the location of the holographic boundary and $\zeta=\zeta_{h}$ refers to horizon determined by $g_{tt}(\zeta_{h})=0$.
The Hawking temperature is given by
\begin{equation}
T=\frac{\left|\kappa\right|}{2\pi},\label{temperature}
\end{equation}
where the surface gravity $\kappa$ is given as
$\kappa=\sqrt{-g_{tt}/g_{\zeta\zeta}}(\text{log}(\sqrt{-g_{tt}}))^{\prime}|_{\zeta=\zeta_{h}}$
\cite{Hashimoto:2016dfz}. Keeping this in hand, it is possible to evaluate the two-point correlators of $J^{\mu}$ by using the partition function of the bulk gravity due to
\begin{equation}
    \left\langle \left[J^{\mu},J^{\nu}\right]\right\rangle =D^{\mu\nu}\propto\frac{\delta^{2}Z_{QFT}}{\delta A_{\mu}^{\left(0\right)}\delta A_{\nu}^{\left(0\right)}}=\frac{\delta^{2}Z_\text{gravity}}{\delta A_{\mu}^{\left(0\right)}\delta A_{\nu}^{\left(0\right)}}.\label{GF}
\end{equation}
As it is known, if QFT is non-perturbative, it would be very challenging to evaluate analytically the correlation function $D^{\mu\nu}$ by using the QFT generating function $Z_\text{QFT}$. However, according to Eq. \ref{GF}, we can achieve this goal by computing the corresponding partition function of the bulk gravity. In this sense, once the bulk action for $A_{\mu}$ as vector meson is given, we can compute the correlation function $D^{\mu\nu}$ of the dual current $J^{\mu}$ by using Eq. \ref{GF}.

Revisiting various models based on AdS/CFT e.g.~\cite{Sakai:2004cn,Sakai:2005yt,Karch:2006pv}, we find while the exact form of the bulk mesonic action depends on the holographic setup we chose, it is possible to write down the quadratic action for vector meson as a generic Maxwell form,
\begin{equation}
S_\text{bulk}=-\int d^{4}x\,d\zeta\,Q(\zeta)\,F_{MN}F^{MN},\label{Action}
\end{equation}
where $F_{MN}=\partial_{M}A_{N}-\partial_{N}A_{M}$ is the gauge field
strength and $M,N$ runs over the five-dimensional bulk. Note that the exact form of $Q\left(\zeta\right)$ depends on the model we chose. Since our concern is the classical partition function $Z_\text{gravity}$ and the correlators $\left\langle \left[J^{\mu},J^{\nu}\right]\right\rangle$ in Eq. \ref{GF}, we need to solve the equation of motion for the bulk vector meson $A_{\mu}$, which can be obtained by varying the action in Eq. \ref{Action} as,
\begin{equation}
\partial_{M}\left[Q(\zeta)F^{MN}\right]=0,\label{EOM}
\end{equation}
then to find its classical solution and $S_\text{bulk}$. By imposing the radial gauge condition $A_{\zeta}=0$ and the Fourier transformation for the remaining components,
\begin{equation}
A_{\mu}(x,\zeta)=\int\frac{d^{4}p}{(2\pi)^{4}}e^{-ip\cdot x}A_{\mu}(p,\zeta),\label{Fourier_trans}
\end{equation}
Eq. \ref{EOM} becomes,
\begin{eqnarray}
 && \partial_{\zeta}\left[Q(\zeta)g^{\zeta\zeta}g^{\mu\nu}\partial_{\zeta}A_{\nu}(p,\zeta)\right]-\nonumber \\
 && \hspace{1cm}p_{\alpha}\left[Q(\zeta)g^{\alpha\beta}g^{\mu\nu}\left(p_{\beta}A_{\nu}(p,\zeta)-p_{\nu}A_{\beta}(p,\zeta)\right)\right]=0,\nonumber \\
 && g^{\mu\nu}p_{\mu}\partial_{\zeta}A_{\nu}(p,\zeta)=0,\label{eq:equations_of_motion}
\end{eqnarray}
where the four-momentum $p_{\mu}$ is given as $p_{\mu}=(-\omega,\bf{p})$. Further taking into account the electric fields associated with the gauge potential
\begin{equation}
E_{i}(p,\zeta)\equiv-p_0 A_{i}(p,\zeta)+p_{i}A_{0}(p,\zeta),
\end{equation}
it leads to the following relations,
\begin{eqnarray} \label{partial_Amu}
\partial_{\zeta}A_{i} & = & -\frac{1}{p_0}\left(\delta^j_i-\frac{p_i p^j}{p^2}\right)\partial_{\zeta}E_{j},\nonumber \\
\partial_{\zeta}A_{t} & = & \frac{p^i}{p_0p^0}\left(\delta^j_i-\frac{p_i p^j}{p^2}\right)\partial_{\zeta}E_{j},
\end{eqnarray}
Hence the first equation in (\ref{eq:equations_of_motion}) reduces to a second order differential equation for $E_i$,
\begin{eqnarray}\label{equations_Ei}
&&\partial_{\zeta}^{2}E_{i}(p,\zeta)+\frac{\left[\partial_\zeta Q(\zeta)g^{\zeta\zeta}\right]}{Q(\zeta)g^{\zeta\zeta}}\left[\partial_\zeta E_i(p,\zeta)\right]-\frac{p^2}{g^{\zeta\zeta}}E_i(p,\zeta) \nonumber \\
&& +(-p_0g_{i\mu}+p_ig_{0\mu})\left(\partial_\zeta g^{\mu\nu}\right)\left[\partial_\zeta A_\nu(p,\zeta)\right]=0,
\end{eqnarray}
where $\partial_\zeta A_\nu$ in the last line can be expressed in terms of $\partial_\zeta E_i$ by using Eq. (\ref{partial_Amu}). Afterwards, we need to find a solution for \ref{eq:equations_of_motion} and \ref{equations_Ei} with incoming wave boundary condition. For the quadratic action \ref{Action} we are considering, the correlation function in \ref{GF} can be written as \cite{Son:2002sd,Casalderrey-Solana:2011dxg},
\begin{equation}
D^{\mu\nu}(p)=\lim_{\zeta\rightarrow0}g^{\zeta\zeta}g^{\mu\alpha}Q(\zeta)\left.\frac{\delta\left[\partial_{\zeta}A_{\alpha}(p,\zeta)\right]}{\delta A_{\nu}(p,\zeta)}\right|_{A_{\mu}(p,0)=0}\,,\label{Green_func}
\end{equation}
where $A_{\mu}(p,\zeta)$ must be determined by solving the equations of
motion in (\ref{eq:equations_of_motion}). Denoting $\widetilde{E}_{i}(e_{j},\zeta)$ as the solution for $E_{i}(p,\zeta)$ satisfying the boundary condition
\begin{equation} \label{boundary_condition}
\lim_{\zeta\rightarrow0}\widetilde{E}_{i}(e_{j},\zeta)=\delta_{ij},
\end{equation}
the correlation function (\ref{Green_func}) can be expressed as
\begin{eqnarray} \label{correlations}
D^{\mu 0}(p)&=&-\lim_{\zeta\rightarrow0}\frac{p_j}{p_0}\left(g^{\mu k}-\frac{p^\mu p^k}{p^2}\right)\partial_\zeta\widetilde{E}_k(e_j,\zeta), \nonumber\\
D^{\mu i}(p)&=&\lim_{\zeta\rightarrow0}\left(g^{\mu k}-\frac{p^\mu p^k}{p^2}\right)\partial_\zeta\widetilde{E}_k(e_i,\zeta).
\end{eqnarray}
Obviously, the Ward identity remains to be $p_\mu D^{\mu\nu}=p_\mu D^{\nu\mu}=0$ with the above discussion. We note that the spectral function for the current-current correlation would come from the imaginary part of $D^{\mu\nu}$,
\begin{equation}
\widetilde{\varrho}_{\lambda\lambda}(p)=v_{\mu}^{\ast}(\lambda,p)v_{\nu}(\lambda,p)\text{Im}\,D^{\mu\nu}(p)\,,\label{spectral_function}
\end{equation}
which is projected onto polarization vectors.

When the correlation function is ready, we will focus on the spin alignment of vector meson. So, for the reader's convenience, we briefly summarize the approach to evaluate the spin alignment of vector meson in the framework of holography as follows,
\begin{enumerate}
\item Solving the equations of motion in (\ref{equations_Ei})
with incoming wave condition at the horizon and condition (\ref{boundary_condition}) at the boundary.
\item Constructing the current-current correlation function using Eq. (\ref{correlations}).
\item Projecting the correlation function onto polarization vectors as shown in (\ref{spectral_function}) and
substituting into Eq. (\ref{eq:dilepton-lambda}) to obtain the dilepton
production rates from different spin states.
\item Averaging over meson's energy and, if necessary, averaging over some specific momentum region to obtain the spin alignment.
\end{enumerate}
In the next section, we will test our method by using QCD soft-wall model and evaluate holographically the spin alignment of vector meson at finite temperature in the strong coupling region.

%%%%%%%%%%%%%%%%%%%%%%%%%%%%%%%%%%%%%%%%%%%%%%%%%%

\section{Holographic spin alignment from QCD soft-wall model} \label{sec:examples}

In this section, we will use the QCD soft-wall model to investigate the spin alignment of vector meson in holography as we discussed in the last section. The holographic background metric in this model is given as,
\begin{equation}
ds^{2}=\frac{L^{2}}{\zeta^{2}}\left(-f(\zeta)dt^{2}+dx^{2}+dy^{2}+dz^{2}+\frac{d\zeta^{2}}{f(\zeta)}\right),\label{eq3}
\end{equation}
where $L$ denotes the AdS radius and $f(\zeta)=1-\zeta^4/\zeta_{h}^{4}$. The Hawking temperature is given by $T=1/(\pi\zeta_{h})$. The action for bulk mesonic field $A_{M}$ takes the form as \ref{Action}, while the function $Q(\zeta)$ is given by
\begin{equation} \label{function-Q}
    Q(\zeta)=e^{-\Phi(\zeta)}\sqrt{-g}/(4g_{5}^{2}),
\end{equation}
where $\Phi(\zeta)$ is the background dilaton field playing the role of infrared cut-off and breaking the conformal symmetry in the soft-wall model. The coupling constant $g_{5}$ can be determined by matching the data of current correlators in the dual QFT (see Refs.\cite{Karch:2006pv,Policastro:2002se,Freedman:1998tz,Grigoryan:2007my}). The simplest choice for $\Phi$ could be $\Phi(\zeta)=c\zeta^{2}$ according to \cite{Karch:2006pv,Mamani:2013ssa,Braga:2015jca,Braga:2016wkm}, since it can reproduce the linear Regge behavior $m_{n}^{2}=4c(n+1)$ for the vector meson mass at zero temperature. While there are some other choices for $\Phi$ \cite{Mamani:2022qnf,Braga:2017bml,Braga:2018zlu,MartinContreras:2021bis}, we will not attempt to use them in this work since they will increase the complexity of our calculation.

\subsection{$J/\psi$ meson} \label{subsec:Jpsi}

We first consider the dimuon decay of the $J/\psi$ meson, which has
the vacuum mass $m_{J/\psi}=3.096$GeV \cite{ParticleDataGroup:2018ovx}. We take the width of $J/\psi$  as $\Gamma\approx100$ MeV instead of its vacuum width, reflecting the phenomenological dimuon production rates observed at LHC \cite{ALICE:2022dyy}. We have checked that varying $\Gamma$ from 100 MeV to its vacuum value does not have a sizable effect on the meson's spin alignment.
For the dilaton field in Eq. \ref{function-Q}, we take $\Phi(\zeta)=c_{J/\psi}\zeta^{2}$
with the parameter $c_{J/\psi}=m_{J/\psi}^{2}/4\approx2.40\text{GeV}^{2}$. When performing the energy integral in Eq. \ref{rho00}, we restrict the invariant mass in the range $2.1 \text{GeV}\leq M\leq 4.9 \text{GeV}$, which is the same as in Ref. \cite{ALICE:2022dyy}.

Considering the $J/\psi$  meson passing through a thermal background, a straightforward choice for the spin quantization direction is $\boldsymbol\epsilon_0=\boldsymbol\epsilon_0^h={\bf p}/|{\bf p}|$, where ${\bf p}$ is the meson's three-momentum in the rest frame of the thermal background. Such a choice corresponds to the so-called helicity frame, because the spatial part of meson's polarization vector for the longitudinally polarized state (state with $\lambda=0$) is parallel to the meson's three-momentum. Due to the motion relative to the background, the symmetry between longitudinally ($\lambda=0$) and transversely ($\lambda=\pm1$) polarized states is broken, leading to different dilepton decay rates $n_\lambda$ and therefore a nontrivial spin alignment.

\begin{figure}[t!]
  \centering
  % Requires \usepackage{graphicx}\label{spectral_1}
  \includegraphics[width=0.4\textwidth]{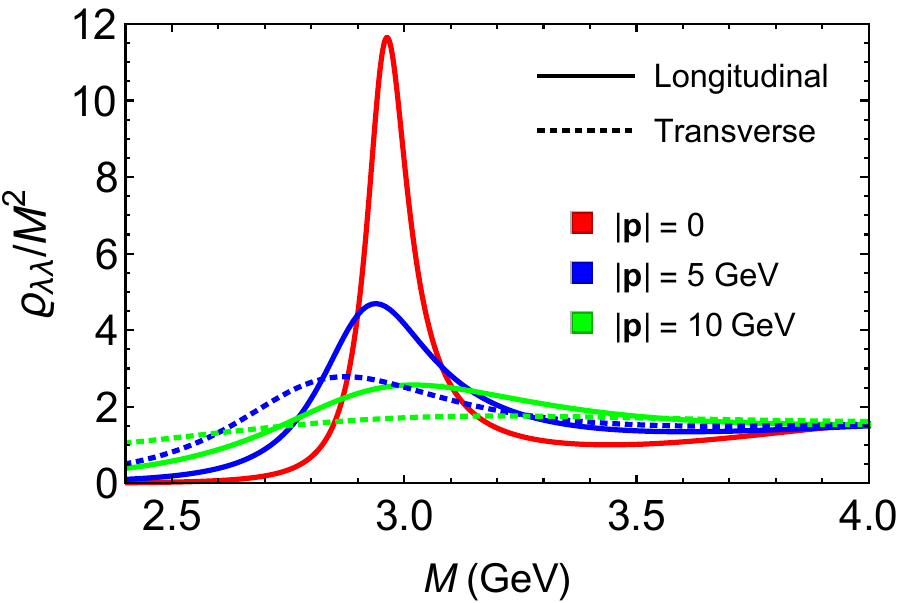}\\
  \caption{Spectral functions $\varrho_{\lambda\lambda}(q)$ as functions of invariant mass $M\equiv\sqrt{\omega^2-{\bf p}^2}$ and momentum $|{\bf p}|=0$ (red lines), 5 GeV (blue lines), and 10 GeV (green lines) at temperature $T=150 \,\text{MeV}$, made to be dimensionless by dividing $M^2$. \label{fig_spectral_function}}
\end{figure}

In Fig. \ref{fig_spectral_function} we plot spectral functions as functions of invariant mass $M\equiv\sqrt{\omega^2-{\bf p}^2}$, computed at temperature $T=150$ MeV. At zero momentum, spectral functions for all spin states are degenerate because of the rotation symmetry, as shown by the red solid line. For mesons with finite momentum, we observe a significant separation between spectral functions for longitudinally polarized state (solid lines) and transversely polarized states (dashed lines). At a small invariant mass $M$, the vector meson is preferred to be transversely polarized. The spectral function for longitudinal polarized state grows and exceeds that for transversely polarized states at $M\sim 2.8$ GeV. We also find that peak values for spectral functions decrease with increasing $|{\bf p}|$, reflecting the fact that high-momentum resonances are harder to be produced than low-momentum resonances.

\begin{figure}[t!]
  \centering
  \includegraphics[width=0.4\textwidth]{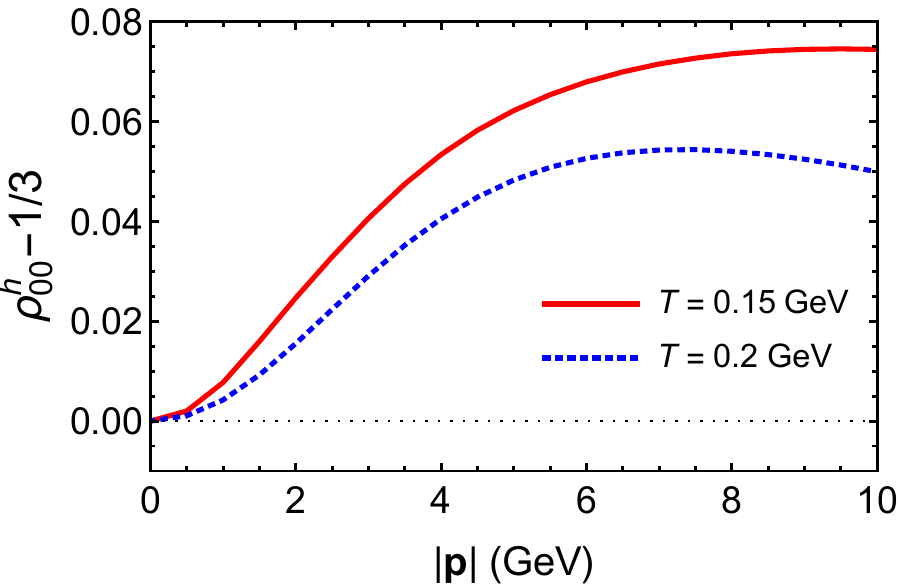}\\
  \caption{\label{rho00-jpsi-helicity} The spin alignment of $J/\psi$ meson in the helicity frame as functions of momentum $|{\bf p}|$ at temperature $T=0.15$ GeV (red solid line) and $T=0.2$ GeV (blue dashed line).}
\end{figure}

With the help of Eqs. (\ref{eq:dilepton-lambda}) and (\ref{rho00}), we are able to calculate the spin alignment (in helicity frame) for $J/\psi$  using spectral functions. The results are shown in Fig. \ref{rho00-jpsi-helicity}, plotted as functions of $|{\bf p}|$ at $T=150$ MeV and $200$ MeV, respectively. We observe a positive deviation from 1/3, corresponding to a negative $\lambda_\theta$ parameter. Such a result is qualitatively different from the experimental results for $J/\psi$  at forward rapidity in Pb-Pb collisions at $\sqrt{s_\text{NN}}=5.02$ TeV \cite{ALICE:2022dyy}. The difference may be because of the $J/\psi$  at forward rapidity region has a much larger momentum than the considered $|{\bf p}|$ in Fig. \ref{rho00-jpsi-helicity}. The deviation from $1/3$ is more significant at lower temperatures.

\begin{figure}[t!]
  \centering
  \includegraphics[width=0.4\textwidth]{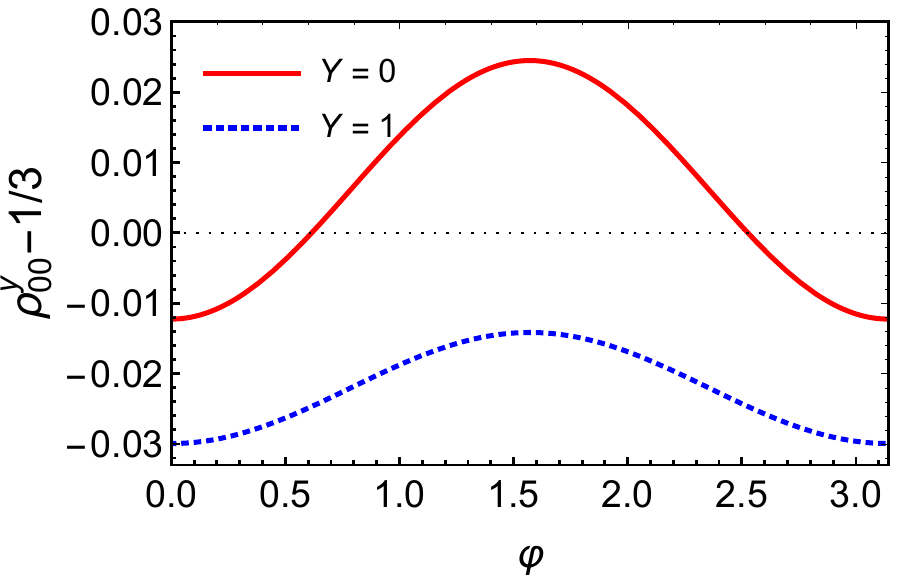}\\
  \caption{\label{rho00-jpsi-theta} The global spin alignment for $J/\psi$ mesons with transverse momentum $p_T=2$ GeV and rapidity $Y=0$ (red solid line) or $Y=1$ (blue dashed line), as functions of the meson's azimuthal angle $\varphi$.}
\end{figure}

\begin{figure}[t!]
  \centering
  \includegraphics[width=0.4\textwidth]{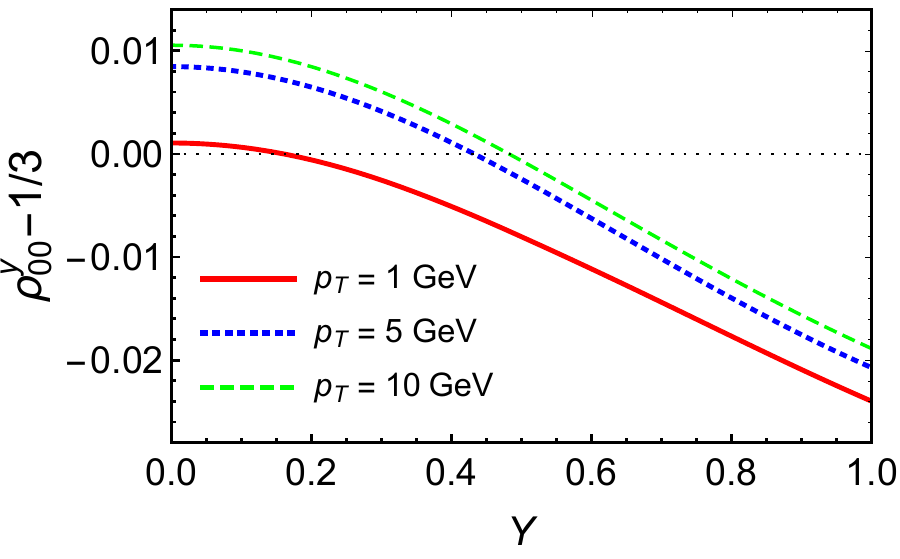}
  \caption{\label{rho00-jpsi-y} The global spin alignment for $J/\psi$ mesons with transverse momentum $p_T=1$ GeV (red line), 5 GeV (blue line), and 10 GeV (green line), as functions of the meson's rapidity $Y$.}
\end{figure}

Now we apply our model to heavy-ion collisions by setting the spin quantization direction as $y$-direction, $\boldsymbol\epsilon_0=\boldsymbol\epsilon_0^y=(0,1,0)$, i.e., the direction of global OAM in heavy-ion collisions, corresponding to the so-called global spin alignment. Using Eq. (\ref{spectral_spin_space-1}) and noting the fact that $\widetilde\varrho_{\lambda\lambda^\prime}$  is diagonal in the helicity frame, we find a relation between diagonal elements of spectral functions for cases with $\boldsymbol{\epsilon}_0=\boldsymbol{\epsilon}_0^h$ and  $\boldsymbol{\epsilon}_0=\boldsymbol{\epsilon}_0^y$,
\begin{eqnarray}
&&\widetilde\varrho_{00}^y(x,p)=\frac{1}{2}\left\{1-\widetilde\varrho_{00}^h(x,p)+(\boldsymbol{\epsilon}_0^y\cdot\boldsymbol{\epsilon}_0^h)^2[3\widetilde\varrho_{00}^h(x,p)-1]\right\},\nonumber\\ &&\sum_\lambda\widetilde\varrho_{\lambda\lambda}^y(x,p)=\sum_\lambda\widetilde\varrho_{\lambda\lambda}^h(x,p)
\end{eqnarray}
where the superscripts $h$ and $y$ labels different choices for $\boldsymbol\epsilon_0$. The global spin alignment is then calculated using Eqs. (\ref{eq:dilepton-lambda}) and (\ref{rho00}). It is straightforward to show that
\begin{equation} \label{relation-rhoy-rhoh}
\rho_{00}^y(x,{\bf p})=\frac{1}{3}+\frac{3\rho_{00}^h(x,{\bf p})-1}{3|{\bf p}|^2}\left(p_y^2-\frac{p_x^2+p_z^2}{2}\right)
\end{equation}
which qualitatively agrees with the prediction of quark coalescence model \cite{Sheng:2023urn}. We plot in Fig. \ref{rho00-jpsi-theta} $\rho_{00}^y$ as functions of meson's azimuthal angle, where the thermal background is assumed to be static while meson's transverse momentum is fixed to $p_T=2$ GeV and the longitudinal momentum is determined by the rapidity $Y$ as $p_z=\sqrt{M^2+p_T^2}\,\sinh(Y)$. We also fix the temperature as $T=150$ MeV. For mesons at center-rapidity $Y=0$, the spin alignment shows a negative deviation from 1/3 when the azimuthal angle $\varphi=0$, then it increases with $\varphi$ and reaches the maximum value at $\varphi=\pi/2$, which is larger than 1/3. At a more forward rapidity $Y=1$, the spin alignment is always smaller than 1/3. The azimuthal angle dependence can be fitted by $\rho_{00}^y-1/3=c_1+c_2 \cos(2\varphi)$, with the parameters $c_1=0.006125$ ($c_1=-0.02205$) and $c_2=-0.01838$  ($c_2=-0.07892$) for $Y=0$ ($Y=1$). By averaging over the azimuthal angle, we obtain in Fig. \ref{rho00-jpsi-y} the global spin alignment as functions of the meson's rapidity, for transverse momentum $p_T=1,\,5,\,10$ GeV, respectively. In order to include the effect of anisotropic flow $v_2$ induced by the initial geometry in collisions, we have performed the average with a weight function $1+2v_2\cos(2\varphi)$ and set $v_2=0.15$. We observed that  $\rho_{00}^y$ increases with increasing $p_T$. For all considered cases, $\rho_{00}^y>1/3$ at center rapidity and decreases to a negative value at a larger rapidity, which is mainly because the difference between the spectral function for the longitudinal mode and that for the transverse modes becomes more significant at larger $|{\bf p}|$. We can naturally expect that $\rho_{00}^y<1/3$ in a more forward rapidity region $2.5<Y<4$, which qualitatively agrees with the ALICE experiment \cite{ALICE:2020iev}.

\begin{comment}
\begin{figure}[t!]
  \centering
  \includegraphics[width=0.32\textwidth]{m-q.pdf}\\
  \caption{ The invariant mass of $J/\psi$ meson as a function of the momentum of the $J/\psi$ meson $q$ for different temperatures.}
\end{figure}
\end{comment}

\subsection{$\phi$ meson}\label{subsec:phi}

Similar to the discussions for the $J/\psi$ meson, we can study the $\phi$ meson within the same framework. Again we apply the dilaton field $\Phi(\zeta)=c_{\phi}\zeta^{2}$ with $c_{\phi}=m_{\phi}^{2}/4\approx0.26\text{GeV}^{2}$ determined by the vacuum mass of $\phi$ meson, $m_{\phi}=1.02\text{GeV}$. We also use the vacuum width $\Gamma=4$ MeV and restrict the $\phi$  meson's invariant mass within the region $1\text{ GeV}\leq M\leq 1.04 \text{ GeV}$. We still focus on the dimuon decay, though it is not the dominant decay channel for the $\phi$ meson, and assume that the spin alignment does not depend on (or weakly depend on) the decay channel. This assumption is reliable because  $\rho_{00}$ can be approximate by $\widetilde{\varrho}_{00}/\sum_\lambda\widetilde{\varrho}_{\lambda\lambda}$ at $p^2=m_\phi^2$,
as demonstrated below Eq. \ref{rho00}, while the spectral function $\widetilde{\varrho}_{\lambda\lambda}$ is independent to the decay process.

\begin{figure}[t!]
  \centering
  \includegraphics[width=0.4\textwidth]{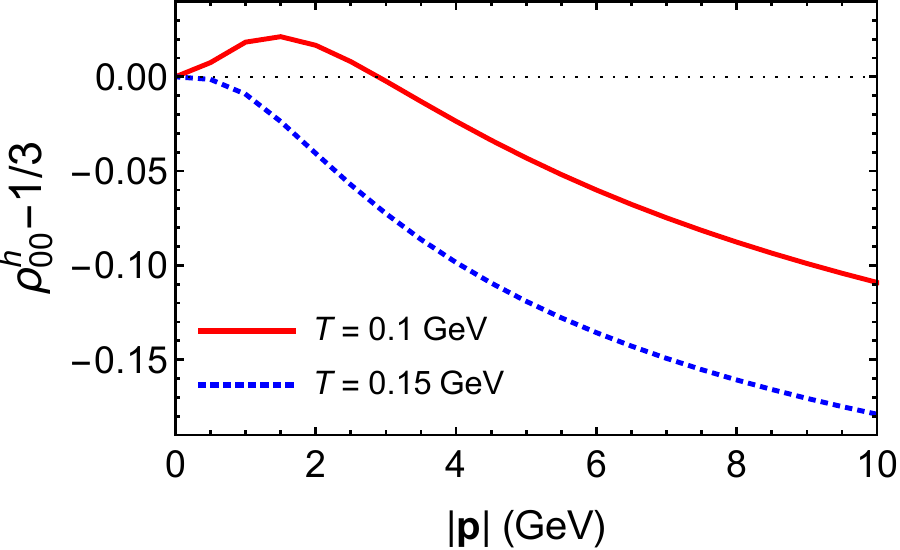}\\
  \caption{\label{rho00-h-phi} The spin alignment for $\phi$ mesons in the helicity frame as functions of momentum $|{\bf p}|$ at temperature $T=0.1$ GeV (red solid line) and $T=0.15$ GeV (blue dashed line).}
\end{figure}

We plot in Fig. \ref{rho00-h-phi} the spin alignment in the helicity frame as functions of $\phi$  mesons' momentum $|{\bf p}|$. For the case of $T=0.15$ GeV, shown by the blue dashed line, $\rho_{00}^h$ shows a negative deviation from $1/3$, whose absolute value becomes larger at larger $|{\bf p}|$. However, $\rho_{00}^h$ at $T=0.1$ GeV (the red solid line) has a non-monotonic dependence to $|{\bf p}|$: it increases with $|{\bf p}|$ in the region $|{\bf p}|<1.5$ GeV, then decreases with $|{\bf p}|$ when $|{\bf p}|>1.5$ GeV. For all considered $|{\bf p}|$, the spin alignment in the helicity frame is larger at $T=0.1$ GeV than that at $T=0.15$ GeV.

\begin{figure}[t!]
  \centering
  \includegraphics[width=0.4\textwidth]{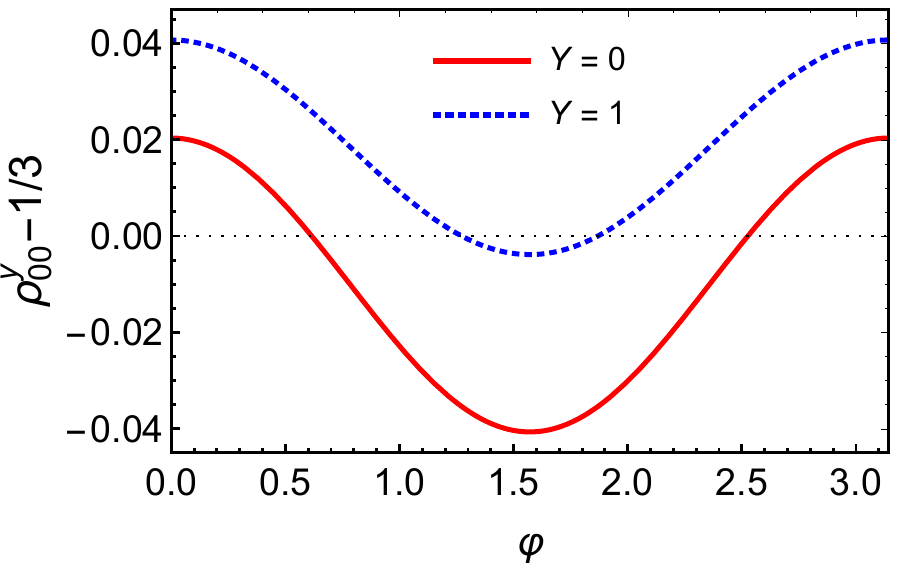}\\
  \caption{\label{rho00-phi} The global spin alignment for $\phi$ mesons with transverse momentum $p_T=2$ GeV and rapidity $Y=0$ (red solid line) or $Y=1$ (blue dashed line), as functions of the meson's azimuthal angle $\varphi$.}
\end{figure}

\begin{figure}[t!]
  \centering
  \includegraphics[width=0.4\textwidth]{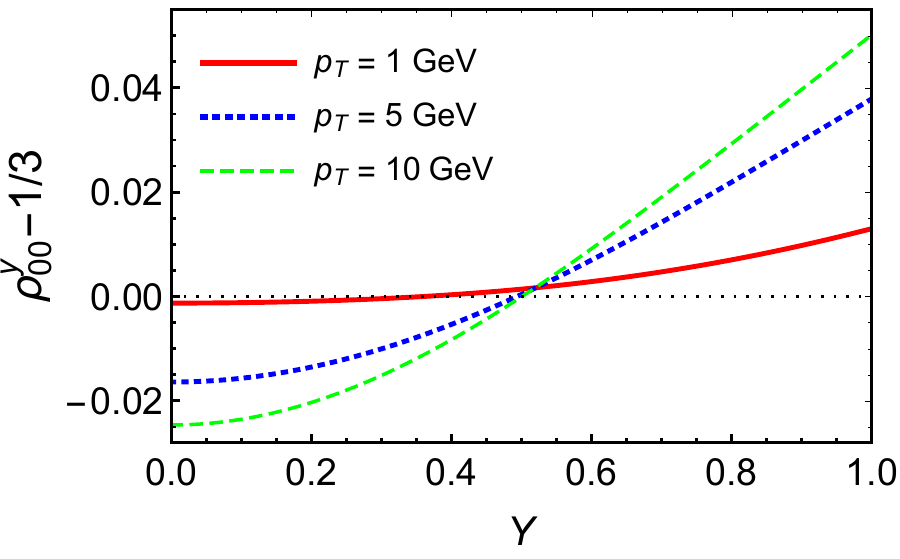}\\
  \caption{\label{rho00-phi-y} The global spin alignment for $\phi$ mesons with transverse momentum $p_T=1 $ GeV (red line), 5 GeV (blue line), and 10 GeV (green line), as functions of the meson's rapidity $Y$.}
\end{figure}

Using relation (\ref{relation-rhoy-rhoh}) between the global spin alignment (measured along the $y$-direction) and the spin alignment in the helicity frame, we are now able to study the global spin alignment for $\phi$ mesons. Fig. \ref{rho00-phi} shows the azimuthal angle dependence of $\rho_{00}^y$ for $\phi$  mesons with a fixed transverse momentum $p_T=2$ GeV at $T=0.15$ GeV.  For mesons at $Y=0$, $\rho_{00}^y$ is larger than 1/3 at $\varphi=0$ and decreases to a minimum value at $\varphi=\pi/2$, which is smaller than 1/3. The results at $Y=1$ has a positive shift compared to results at $Y=0$. Then we average over the azimuthal angle with a weight function $1+2v_2\cos(2\varphi)$ with $v_2=0.15$ and plot in Fig. \ref{rho00-phi-y} $\rho_{00}^y$  as functions of rapidity $Y$ at $p_T=1$ GeV, 5 GeV, and 10 GeV, respectively. For all three cases, we find $\rho_{00}^y<1/3$ when $Y<0.5$,  and it increases to a positive value when $Y>0.5$. Absolute values for deviations from 1/3 become larger at larger $p_T$, whose magnitude can be as large as $\mathcal{O}(10^{-2})$, which is the same magnitude as those observed in STAR experiments \cite{STAR:2022fan}.

The global spin alignment for $\phi$  mesons obtained in this paper quantitatively agrees with the prediction by quark coalescence model at 200 GeV \cite{Sheng:2023urn}. However, Ref. \cite{Sheng:2023urn} has used parameters extracted from experiment measurements for global spin alignment, while calculations in this paper do not rely on any subjective parameter except the meson's physical vacuum mass and vacuum width.  Therefore the holographic approach may provide a possible way to study spin alignment from first-principle calculations.

\section{Summary} \label{sec:summary}

We have studied the spin alignment of flavorless vector mesons with a holographic description. For dilepton decays of vector mesons, we have derived a relation between the production rate and the imaginary part of the current-current correlation function, which is the spectral function of vector meson. According to AdS/CFT dictionary, the correlation function in QFT can be perturbatively calculated in a dual gravity theory on the AdS space. We have accordingly derived the equations of motion of the bulk vector meson field, which are expressed in terms of the corresponding electric fields. The correlation function have been then calculated by boundary values of fields and their derivatives with respect to the holographic radial coordinate,
enforcing the incoming wave boundary conditions at the horizon of the bulk.

As an example of the holographic description, we have considered the soft-wall model in Sec. \ref{sec:examples}, for both $J/\psi$  and $\phi$  mesons. Without incorporating external fields like magnetic or vorticity fields, the spin alignment is purely induced by the motion of vector meson relative to the background. As a consequence, the spin alignment in the helicity frame $\rho_{00}^h$ deviates from 1/3 when the meson has a non-vanishing momentum. Surprisingly, the model prediction shows that $J/\psi$ and $\phi$ have opposite behaviours. At a fixed temperature $T=150$ MeV, $\rho_{00}^h>1/3$ for $J/\psi$, while $\rho_{00}^h<1/3$ for $\phi$ mesons, indicating that $J/\psi$  ($\phi$) mesons are more likely to be longitudinally (transversely) polarized. This difference may arise from their different spectral functions because of different masses. We also study their global spin alignment $\rho_{00}^y$, i.e., the spin alignment measured along the direction of global OAM in heavy-ion collisions. For both $J/\psi$ and $\phi$, the azimuthal angle dependence can be fitted to an analytical form $\rho_{00}^y-1/3=c_1+c_2 \cos(2\varphi)$. For $J/\psi$ mesons, the parameter $c_1$ decreases with increasing rapidity $Y$ and reaches a negative value at $Y=1$, while for $\phi$ mesons the behaviour is opposite. One can further expect that, in a more forward rapidity region $2.5<Y<4$, $\rho_{00}^y$ for $J/\psi$ could also be smaller than 1/3 corresponding to $\lambda_\theta>0$, which would agree qualitatively with the recent ALICE experiment observations \cite{ALICE:2022dyy}. This rapidity region is beyond the scope of this
paper because the related momentum is too large.

We emphasize that the global spin alignment for $\phi$ mesons given by the soft-wall model in this paper quantitatively agrees with a previous prediction obtained within the quark coalescence model \cite{Sheng:2023urn}. In Ref. \cite{Sheng:2023urn}, the spin alignment is explained
in terms of fluctuations of the strong force field, with the rotation symmetry being broken by the motion of vector meson and the magnitude of fluctuation being extracted by fitting experiment results. The consistency between two models may imply that the meson's spin alignment is a non-perturbative property in the strongly interacting matter. The validity of our model predictions for the azimuthal angle dependence and rapidity dependence of  $\rho_{00}^y$ for $J/\psi$ and $\phi$  mesons in mid-rapidity region00 $|Y|<1$  would be tested in future experiments.

\section*{Acknowledgement}
The authors thank Hai-Cang Ren for enlightening discussions. This work is supported in part by the National Key Research and Development Program of China under Contract No. 2022YFA1604900. This work is also partly supported by the National Natural Science Foundation of China (NSFC) under Grants No. 12275104,  No. 11735007. Si-wen Li is supported by the National Natural Science Foundation of China (NSFC) under Grant No. 12005033 and the Fundamental Research Funds for the Central Universities under Grant No. 3132023198.

\bibliographystyle{apsrev}
\bibliography{Holographic-general}

\end{document}